\documentclass[aps,prl,twocolumn,superscriptaddress,nofootinbib]{revtex4-1}
\usepackage[utf8x]{inputenc}
\usepackage{amsmath}
\usepackage[dvips]{graphicx}

\usepackage{color}

\begin{document}


\title{Implications of the UHECRs penetration depth measurements}


\author{Nimrod Shaham}
\author{Tsvi Piran}

\affiliation{The Racah institute of Physics, The Hebrew university of Jerusalem}

\date{\today}

\begin{abstract}
The simple interpretation  of PAO's  UHECRs' penetration depth   measurements  suggests a  transition at the energy range $1.1 - 35  \cdot 10^{18} $ eV from protons to heavier nuclei.  A detailed comparison of this data with air shower simulations reveals strong restrictions on the amount of light nuclei (protons and He) in the observed flux.
We find a robust upper bound on the observed proton fraction of the UHECRs flux and we rule out a  composition dominated by protons and He. Acceleration and propagation effects lead to an observed composition that is different from the one at the source.  Using  a simple toy model that take into account these effects, we show that the observations requires an extreme metallicity at the sources  with metals to protons mass ratio of 1:1, a ratio that is larger by a factor of a hundred than the solar abundance. 
This composition imposes an almost impossible constraint on all current astrophysical models for UHECRs accelerators.  This may provide a first hint towards new physics that emerges at $\sim 100$ TeV and leads to a larger   proton cross section at these energies. 
\end{abstract}


\maketitle

Among the most interesting  results of the Pierre Auger Observatory (PAO) are  the penetration depth measurements \cite{Xmax1,Xmax2,PAOprlc}: the observed mean depth where maximal number of secondaries are generated, $\langle X_{max} \rangle(E)$, and the fluctuations of this quantity $RMS(X_{max})(E)\equiv \sigma(X_{max}))$.  
When compared with extensive shower simulations
 \cite{QGSJET,QGSJET01,EPOS,Sibyll}   
 both  $\langle X_{max} \rangle$ and $\sigma(X_{max})$ are consistent with a protonic composition around $1~EeV$.  However, at higher energies,   $\langle X_{max} \rangle$ and $\sigma(X_{max})$ fall below the protonic simulated values, towards values estimates for intermediate  mass (e.g. iron) nuclei\footnote{The interactions between UHECRs and atmospheric particles takes place  at  CM energies of  $\sim100$ TeV. This is  about  a hundred times higher than energies in which cross sections have been measured.} .  At first sight this suggests a transition, around $10$  EeV,  from protons to intermediate mass  nuclei \cite{Xmax1}\footnote{While PAO has the best statistics, the High resolution fly's eye (HiRes) and the telescope array (TA) observatories \cite{TA, HiRes} composition results  are consistent with proton dominated composition all the way up to the high end of the spectrum.   This disagreement among the different observatories is still to be settled.}.  We explore here the implications of the penetration depth data combined with the observed energy spectrum \cite{spec1,spec2,PAOprlf} .


Consider  a cosmic ray flux composed of N species each with a fraction $f_j$,  a mean penetration depth, $\langle X_{max} \rangle_j$, and RMS variation $\sigma_j $. At each energy these are  related to the measured mean and RMS values as:
\begin{equation}
 \sum \limits_{j=1}^N f_j  \langle X_{max} \rangle_j = \langle X_{max} \rangle  , \label{meantot}
\end{equation}
 \begin{equation}
 \sum \limits_{j=1}^N f_j  (\sigma_j^2+\langle X_{max} \rangle_j^2) - \langle X_{max} \rangle^2 = \sigma(X_{max})^2  . \label{rmstot}
\end{equation}
In the following we examine several possible solutions of these equations for different compositions.   

Consider, first,   a mixture of protons (denoted p) and another arbitrary component, denoted $0$.   
This arbitrary component may be a single species or a combination of a few. 
The condition  $ \sigma_0^2>0$ yields an upper bound on $f_p(E)$:
\begin{eqnarray}
\label{upper}
& f_p \le  \frac{1}{2} \biggl\{1+ \frac{\sigma(X_{max})^2+(\langle X_{max} \rangle -\langle X_{max} \rangle_p )^2} {\sigma_p^2} - \\
\nonumber
&\sqrt{ 
\bigl[1+\frac{\sigma(X_{max})^2+(\langle X_{max} \rangle -\langle X_{max} \rangle_p )^2}{\sigma_p^2}\bigr]^2 -\frac{4\sigma(X_{max})^2}{\sigma_p^2} }~ \biggr\} .
\end{eqnarray}

\begin{figure}
\includegraphics[width=3in,height=2in]{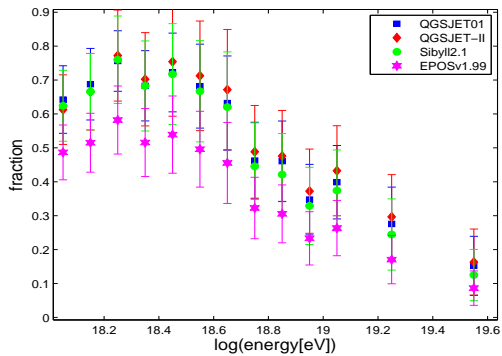} 
\caption{An upper bound on the proton fraction of the observed flux as a function of energy, calculated using the  PAO data and different extensive air shower simulations: QGSJET01 \cite{QGSJET01},  QGSJET-II \cite{QGSJET} , Sibyll2.1 \cite{Sibyll}
and EPOSv1.99 \cite{EPOS} (see legend).}
\label{fig:b2}
\end{figure}

Fig. \ref{fig:b2} depicts the upper limit obtained using the observed  values of $\langle X_{max} \rangle$  and $\sigma(X_{max})$ and the simulated  values  of $\langle X_{max} \rangle_p$ and $\sigma_p$. 
The maximal proton fraction is smaller than  $50 \%$ at  $E>10^{19}$ $eV$  and it decreases below 30\% at higher energies. 
This upper bound  depends only on the observed PAO  data and on the shower simulation results for protons (it does not even depend on the shower simulations for nuclei).  
It is  independent  of the acceleration or propagation of the UHECRs.  
When we replace the hypothetical ingredient with any  composition of He, N, Si and Fe 
the  upper bound on $f_p$ at the two highest energy bins  is smaller by $\approx  10\%$ than the upper limit derived using $\sigma_0=0$.

Protons and Fe survives best the interactions with the cosmic photon field while propagating from the sources to Earth \cite{olinto2010review}. As such they  are the most natural UHECR ingredients. However, 
with just  two components the  system of eqs. \eqref{meantot}-\eqref{rmstot} is overdetermined.
Wilk  and W{\l}odarczyk  \cite{Wilk2010} have shown that  there is no consistent solution for protons and Fe within the error bars of the PAO data and the shower simulations. 

\begin{figure}
\centering
\includegraphics[width=3in,height=2in]{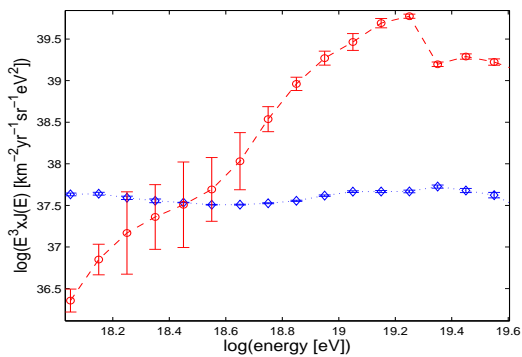}
\caption{The total observed total flux (blue dotted line with diamonds) and the calculated secondary proton flux arising from He disintegration (eq. \eqref{pHeflux}) (red dashed line with circles), for a composition with a maximal number of primary protons 
(satisfying eq. \eqref{upper})
and He (color online). At high energies the  number of secondary protons exceeds the observed flux. }
\label{fig:pHe}
\end{figure}

Eqs.  \eqref{meantot}-\eqref{rmstot} have a marginal ($\chi^2 / dof \approx 1$) solution for a 
mixture of protons and He (see also \cite{unger2012}). However,  when propagation effects are taken into account this composition
can be ruled out.   He nuclei photodisintegrate rapidly on their way from the sources to Earth.   For simplicity we neglect redshift effects and assuming a uniform distribution of sources.   Under this approximations the fraction of He nuclei with energy E surviving and reaching earth is  $F_{{GZK}_{He}}(E)= l_{HE}/l_h \ll 1$,  where $l_{HE}$ is   the  mean free path for He photodisintegration  and $l_h$ is 
the horizon size. 
For a given observed  He flux, $(1-f_p(E)) J(E)$, the flux at the source is  $(1-f_p(E)) J(E)/F_{{GZK}_{He}}(E)$. Since $F_{{GZK}_{He}}(E)\ll1 $ most of the He nuclei  disintegrate producing secondary protons with energy $E/4$. The  resulting secondary proton flux,  $J_{p,sec}(E)$ is\footnote{(i) The protons GZK distance is comparable to the horizon distance at these energies. (ii) We have overestimated here the GZK distances of $^3He$ and $^2H$ as equal to the GZK distance of $^4He$.}:
\begin{equation}
 J_{p,sec}(E) \approx 4(1- F_{{GZK}_{He}}(4E))^3 \frac{(1-f_p(4E)) J(4E)}{ F_{{GZK}_{He}}(4E)}
 \label{pHeflux}
 \end{equation} 
Using the upper bound on $f_p(E)$ obtained earlier (eq. \eqref{upper}) we find (see fig. \ref{fig:pHe}) that this secondary proton flux  is larger than the maximal proton flux allowed.

Kampert and Unger \cite{unger2012} have shown that a  mixture of protons, He and intermediate  elements like N, Si and Fe  can provide a solution of   eqs.  \eqref{meantot} and \eqref{rmstot} for the observed composition which is within the uncertainties of the observed data and the simulations.  
However, as we have seen for p and He, to determine the sources' composition we need to take propagation effects into account. Moreover, since different species are accelerated differently within a given accelerator,  acceleration should also be considered. 
To examine these effects  we consider a toy model based on only  two components: protons and Fe.  
 As mentioned earlier, this composition cannot satisfy equations \eqref{meantot} and \eqref{rmstot} and other intermediate elements in addition to Fe are needed.  Therefore, when we examine a proton and Fe composition we consider only    the overall spectra and $\langle X_{max} \rangle$ and we ignore, for simplicity, the RMS data.  This simple example  is sufficient for demonstrating  the nature of the problem.

Any electromagnetic acceleration process that accelerates protons to 
energy $E$ accelerates nuclei  (with charge $Z$) to  energy \footnote{In the diffusive shock acceleration, a proton and a nucleus, with an atomic weight A,  crossing the shock front  will have the same Lorentz factor, and therefore a nucleus will be A times more energetic. This suggest that we  need to compare  nuclei at energy  E with protons with energy  $E/A$  and not $E/Z$. This will add a factor of  $ (A/Z)^{\alpha-1} \sim 3 $ to the composition ratio. This factor doesn't change qualitatively the results}  $Z E$.   
This suggests a natural \cite{Allard2011,Aloisio:2011ev}  explanation for the transition in composition: the source accelerates protons to a power law energy distribution, $E^{-\alpha}$ up to some maximal energy where a gradual cutoff begins. 
The same source accelerates  nuclei to the same power law  but up to an energy that is $Z$ times larger.  This naturally produces a heavier observed composition than the one at the  source and may suggest that 
the drop at very high energies in the UHECR flux is not necessarily due to a GZK effect but simply due to lack of available 
 accelerators that can accelerate UHECRs to extremely large energies \cite{disappointing}.   We characterized  the accelerator's cutoff by an unknown function of the rigidity, $g({E}/{Z})$, with $0 \leq g(E) \leq 1$ and 
$g(E)=1$ at  low energies. 

However, before  adopting this model propagation effects should also be taken into account. Like acceleration,  propagation in the IGM magnetic field depends on the rigidity. On the other hand GZK attenuation depends on the nucleus at hand. In this energy range { (1.1-35 EeV)}, it is negligible for protons but it is significant for all nuclei.
We characterize the propagation effects using $F_{{GZK}_{Fe}}=l_{Fe}/l_h$ \cite{Allard2011}.  Under these assumptions  the total observed UHECRs flux is:
 \begin{equation}
J(E)=c_p g(E)  E^{-\alpha} 
+c_{Fe}  F_{{GZK}_{Fe}}(E)  g(\frac{E}{26})  \frac{E^{-\alpha}}{26^{1-\alpha}}  
\label{a1}
\end{equation}
{Where 
$c_{p},c_{Fe}$ are normalization factors for the proton and  Fe nuclei fluxes respectively. $c_p$ is obtained by the condition $g(E)=1$ at the minimal energy. The normalization of $c_{Fe}$  is such that $c_{Fe}/c_{p}$ equals the 
Fe nuclei to  protons number ratio at the source (before acceleration)\footnote{Within this acceleration model 
the number of  Fe nuclei accelerated to energies larger than 26E equals to $c_{Fe}/c_{p}$ times the
number of protons accelerated to energies larger than E  (Note that   \cite{ThePierreAugerCollaboration2011} use somewhat different notations)}. }
Using the proton fraction $ f_{p}(E) \equiv {c_p  g(E)  E^{-\alpha}}/{J(E)}$ and $g(E/26)=1$, which is valid  over the relevant energy range { (1.1-35 EeV)}
eq. \eqref{a1} becomes:
\begin{equation}
 \frac{J(E)(1-f_{p}(E))}{F_{{GZK}_{Fe}}(E)} =  
  \frac{c_{Fe}}{26^{1-\alpha}} E^{-\alpha} ,
 \label{a4} 
\end{equation}
{We solve  eq. \eqref{meantot} for  $f_p(E)$    using  the measured $\langle X_{max} \rangle$ and the simulated values for protons and Fe.  Now that all the quantities at the l.h.s of \eqref{a4} are known we fit a power law ($E^{-\alpha}$) to  the l.h.s to  to obtain $\alpha$
and $c_{Fe}$.  The best fit results are: $\alpha = 2.1\pm 0.1$ ,   $c_{Fe}/c_p = (2.2 \pm 0.6) \cdot 10^{-2} $, with  ${\chi^2}/{dof}=0.21$.}

 \begin{figure}[h]
\centering
\includegraphics[width=3in,height=2in]{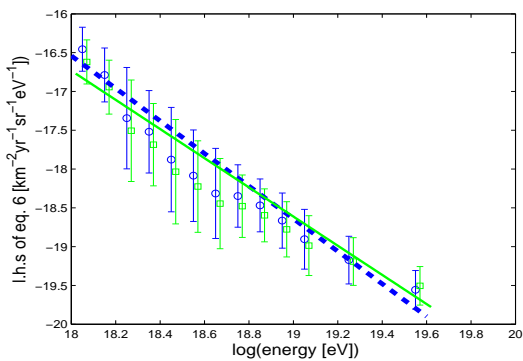} 
\caption{ The l.h.s of eq. \eqref{a4} ( with errorbars, blue circles for Fe, green squares for Si) and the best fit for the r.h.s  of eq. \eqref{a4} (green solid line for Si, blue dashed line for Fe, color online). The Si results are shifted 
to the right for clarity. The best fit results for Fe and protons: $\alpha = 2.1\pm 0.1$ ,  $c_{Fe}/ c_p = (2.2 \pm 0.6) \cdot 10^{-2} $, with  ${\chi^2}/{dof}=0.21$ and for Si and protons: 
$\alpha = 1.9 \pm 0.1$, $c_{Si}/ c_p = (5 \pm 1) \cdot 10^{-2} $, with ${\chi^2}/{dof}=0.4$. }
\label{fig:powermean}
\end{figure}

The number ratio we obtained corresponds to a mass ratio of  $ \approx 1:1$. A similar analysis with 
 Si instead of Fe yields $\alpha = 1.9 \pm 0.1$, $c_{Si}/ c_p(E) = (5 \pm 1) \cdot 10^{-2} $, with ${\chi^2}/{dof}=0.4$. 
 The mass ratio is again $ \approx 1:1$.
As both Fe and Si show almost the same trend  a composite solution that will satisfy both the  $\langle X_{max}\rangle$ and 
$\sigma(X_{max})$ data \cite{unger2012} will have similar features and a  $\sim 1:1$ mass ratio between the protons and the metals. 

Note that we have neglected  secondary protons arising from  photodisintegration. Taking those into account would have resulted in even higher metallicity. 
Shifting the observed  energies by about 20\%, as suggested by comparison of the PAO spectra with the spectra observed in other  
main UHECRs observatories \cite{compare},  does not change qualitatively our results. 
Interestingly, this extremely heavy composition at the source   is comparable to the upper limit obtained  
using the angular distribution 
\cite{ThePierreAugerCollaboration2011} of these UHECRs. These ratios of ${N_{Fe}(>26E)} / {N_p(>E)}$ are $>0.072$
for the  VCV catalogue and  $ >0.084$ for  correlations with Cen A.

Finally, we note that the spectral index, found here,  is much harder than what observed in lower energies, $\alpha=3$ \cite{flyseye}. This arises from  the GZK attenuation affecting nuclei at these energies. 
Such hard spectra were obtained in detailed propagation simulations  \cite{Allard2011,Aloisio:2011ev,Hooper2010}.

 We have shown that the PAO penetration depth measurements and the  penetration depth numerical simulations yield a robust upper limit  on the observed proton fraction of the UHECRs flux. This limit  drops below $50\%$   at energies higher than $10$ EeV and below 30\% at higher energies.   These measurements are inconsistent  with the  composition  of 75\%  protons and 24\% He, that is common in the Universe and they require a significant fraction of intermediate mass nuclei. 

The conversion of the observed composition to the composition at the source depends on the acceleration and  propagation.  Using  a simple toy model we find  that  the protons to metals mass ratio at the source should be about $1:1$.
This metallicity is larger by a factor of a hundred than the solar  metal abundance of $\approx 1$\%, which   reflects typical metallicity in the Universe. 
This high metallically  puts  a new  severe constraint on the sources, since objects  dominated by nuclei heavier than He are rare in the astrophysical landscape. 


Active galactic nuclei (AGNs) are  natural  UHECRs accelerators  \cite{cavallo1978,Romero1996}.
Most AGNs  can accelerate metals to $\sim 10^{20}$eV and protons to an order of magnitude lower. This would produce, naturally, the observed composition transition 
\cite{Allard2011,Aloisio:2011ev} as well as the cutoff in the spectrum at higher energies \cite{disappointing}.
However this would  require a very heavy composition, whereas  AGNs typically show  solar-like metallicities \cite{abundancesreview}.

Gamma ray bursts (GRBs) are another natural  UHECRs accelerators \cite{Waxman95} (see however \cite{Guetta2010,IceCube2012}). If  UHECRs are produced by GRBs' internal shocks, they are composed of the original material of the 
jets.  One has to invoke, in this case, a very efficient nucleosynthesis within the jets and survival of the produced nuclei during the acceleration \cite{Metzger2011,Shunsaku2012}.   It is not clear that this can be achieved generically. Recall that the UHECR  output of GRBs should be comparable or larger than  their $\gamma$-rays output. Alternatively if UHECRs are produced by external shocks they will be composed by the circum-burst winds surrounding the star. GRBs' progenitors, Wolf - Rayet stars,  have He, C and O dominated wind whose composition is too light.   

A variation on this theme was proposed by Liu and Wang \cite{Ruoyu2012} who  invoke, instead of regular GRBs,  low-luminosity GRBs which they describe as ``semi relativistic core collapse supernovae that involve an engine activity" and  call ``hypernovae". These are also based on Wolf - Rayet progenitors. However,  low-luminosity GRBs  jets do not penetrate the stellar envelope and the observed emission is produced by a shock breakout \cite{Bromberg2012}.  
Thus, it is not clear how could these bursts accelerate UHECRs in the first place. Furthermore,  as sources of heavy UHECRs they suffer from all problems mentioned earlier concerning regular GRBs.  


A rapidly rotating young pulsar   with a strong but reasonable magnetic field ($\sim 10^{13}$ G) can accelerate Fe to UHECR energies \cite{Arons2003}.  Fang et al. \cite{Fang2012} suggested that young pulsars are UHECR sources and the origin 
of the heavy composition is the Fe rich crust.   However, X-ray observations of neutron stars suggest the existence of an atmosphere composed of proton and light elements  above the crust (see \cite{Arons2003} and citations therein).  Acceleration of this  atmospheric  component will result in a  light  composition. This poses a serious doubt concerning this model.


Overall, while astrophysical heavy UHECRs sources  cannot be ruled out with absolute certainty, the strict constraint on the composition obtained here (that constrain both the protonic and the He components) makes the UHECRs sources puzzle even harder to solve. In particular it rules out the most natural  sources, AGNs.  There is no single clear model that naturally produces UHECRs with such a composition. { One may search for acceleration processes that are not rigidity dependent and favor heavy nuclei over light ones, however such  a process is not readily available. }

Given this situation 
one can consider the following alternatives to the heavy composition.
First, the observational data might be incorrect or  somehow dominated by poor  statistics: these 
results are  based on  about  1500  events at the lowest energy bin and on only about 50 at the highest one.
The possibility of a  miscalculation in the shower simulations is unlikely. Different simulations \cite{QGSJET,QGSJET01,EPOS,Sibyll}  obtain comparable results. 
However, the simulations depend on the extrapolations of the  proton's cross section from the measured energies of a few TeV to the range of an UHECR - atmospheric nuclei collision, which are factor of 100 higher in energy.   Is  it possible that this extrapolation breaks down?  A larger cross section than the one extrapolated  can explain the shorter penetration depth.
If so these findings might provide  hints of  a new physics that set in at  energies of several dozen TeV\cite{Farrar}.

This research was partially supported by an ERC Advanced research grant and by the ISF center for High Energy Astrophysics. 



%

\end{document}